\def\CommentsEnabled{}

\documentclass[journal, comsoc]{IEEEtran}

\usepackage{cite}

\ifCLASSINFOpdf
\else
\fi

\usepackage{amsmath}
\usepackage{algorithmic}

\usepackage{url}
\usepackage{float}
\usepackage{graphicx}
\usepackage{color}

\ifx\CommentsEnabled\defined
	\newcommand{\massimo}[1]{}%
	\newcommand{\mattia}[1]{}%
\else
	\newcommand{\massimo}[1]{\textcolor{red}{(#1)}}
	\newcommand{\mattia}[1]{\textcolor{blue}{(#1)}}
\fi

\newcommand{\ie}{\textit{i.e.}}
\newcommand{\eg}{\textit{e.g.}}

\newcommand{\myfig}[1]{Figure~\ref{#1}}

\newcommand{\mysec}[1]{Section~\ref{#1}}

\usepackage{epstopdf}
\DeclareGraphicsExtensions{.eps}
\graphicspath{{./FIGURES/PDF/},{./FIGURES/EPS/}}

\usepackage{hyperref}

\begin{document}

\title{Fog Computing Architectures:\\a Reference for Practitioners}
%
%
%

\author{Mattia~Antonini,
        Massimo~Vecchio,
        and~Fabio~Antonelli
\thanks{M. Antonini, M. Vecchio and F. Antonelli are with the OpenIoT research unit, FBK CREATE-NET, Trento, TN, 38123 Italy e-mail: m.antonini@fbk.eu, mvecchio@fbk.eu and fantonelli@fbk.eu.}
\thanks{M. Antonini is also with the Department of Information Engineering and Computer Science, University of Trento, Trento, TN, 38123 Italy email: mattia.antonini@unitn.it}
\thanks{Accepted for Publication in IEEE Internet of Things Magazine. \copyright~2019 IEEE. Personal use of this material is permitted. Permission from IEEE must be obtained for all other uses, in any current or future media, including reprinting/republishing this material for advertising or promotional purposes, creating new collective works, for resale or redistribution to servers or lists, or reuse of any copyrighted component of this work in other works.}}

%
%

\markboth{IEEE Internet of Things Magazine}%
{Antonini, Vecchio, Antonelli: Fog Computing Architectures: a Reference for Practitioners}
%



\maketitle

\begin{abstract}
Soon after realizing that Cloud Computing could indeed help several industries overcome classical product-centric approaches in favor of more affordable service-oriented business models, we are witnessing the rise of a new disruptive computing paradigm, namely Fog Computing. Essentially, Fog Computing can be considered as an evolution of Cloud Computing, in the sense that the former extends the latter to the edge of the network (that is, where the connected devices --the things-- are) without discontinuity, realizing the so-called \textit{``cloud-to-thing continuum''}. Since its infancy, Fog Computing has been considered as a necessity within several Internet of Things (IoT) domains (one for all: Industrial IoT) and, more generally, wherever embedded artificial intelligence and/or more advanced distributed capabilities were required. Fog Computing cannot be considered only a fancy buzzword: according to separate, authoritative analyses its global market will reach \$18 billion by 2022, while nearly 45\% of the world's data will be moved to the network edge by 2025. In this paper, we take stock of the situation, summarizing the most modern and mature Fog Computing initiatives from standardization, commercial, and open-source communities perspectives.
\end{abstract}


%

\section{Introduction}
\label{sec:intro}
The original mission of the Internet of Things (IoT) was to connect devices to the Internet, enabling communications and autonomous interactions among everyday objects. Since the beginning of the IoT era, sensing, actuation, and communications have been the main tasks of the designers and the integrators of these devices. As a matter of facts, always tinier and more powerful devices are reaching the global market at very affordable prices. However, in several modern vertical domains, applications require always more computational capabilities than those available on commercially available ``smart'' objects. In such cases, the classical approach is to send data to remote cloud endpoints that have, ideally, infinite computational power and storage capabilities. Here it is possible to run computational intensive operations like machine learning tasks, data aggregation, storage, monitoring, and visualization. In this respect, many researchers and big companies have developed solutions and products able to reliably receive and remotely process data coming from multitudes of IoT devices. This approach has enabled a new class of applications able to leverage on data coming from geographically distributed deployments of sensing and actuators devices. However, this approach, often referred to as Cloud-centric, suffers from many problems related to the communication between devices and cloud end-points. Indeed, the delay between a device's request and the corresponding cloud's response may cause unpredictable effects on the surrounding environment, especially in latency-sensitive applications. Furthermore, such a system needs an always-on connection to work properly. This means that a failure may reduce the reliability of the whole application. Sensitive data should not be transmitted and processed in remote systems since they can be eavesdropped, corrupted or even destroyed. Last but not least, connected devices may produce volumes of data that cannot be transmitted-to/handled-by remote systems. These are, for instance, the typical challenges that an Industrial IoT scenario has to face.\\\\
A possible solution to alleviate these issues was proposed by Bonomi in 2012 when he first introduced the Fog Computing paradigm as an extension of the Cloud Computing capabilities to the edge of the network \cite{Bonomi2012}. Fog Computing aims to move the execution of tasks from the Cloud closer to the data sources, by exploiting networking entities like gateways, access points, and routers. The Fog Computing architecture that he originally proposed comprised three main layers: the bottom layer, containing IoT devices able to sense and act in the surrounding environment; a middle layer, that is the Fog layer, responsible for processing data locally and, if needed, able to forward them to a remote cloud system; and the upper layer, that is the Cloud itself. Fog Computing introduces the concept of Cloud-to-Thing continuum since the data processing from the sensing device to a cloud endpoint can be distributed along the way without discontinuity. This paradigm addresses, with a by-design approach, some of the major issues of the today's cloud-based solutions with respect to privacy, reliability, latency, and bandwidth, at the additional cost of increasing the local computing power. This approach has the potential to enable new classes of applications; just as an example, consider a condition monitoring application, where some dedicated sensors sample the current state of an engine and a software application takes suitable actions based on the data acquired from the field. Here, we can have two alternative approaches: a cloud-based and a fog-based approach. In the first case, we deploy some sensors in the field that collect information about the physical phenomenon and forward it to a cloud endpoint, in charge of executing the required processing. If an anomalous condition is met, the cloud application sends a command to some actuators to take an action on the environment (\eg, halting the monitored engine). In this case, the bandwidth required to stream the data might be too large, or the latency too high to react sufficiently fast to avoid a dangerous event. Moreover, if the connectivity fails and an anomalous event happens, then the end-to-end application might not be able to react, causing even serious consequences. The orthogonal possibility is to develop an application leveraging the Fog Computing paradigm. In this case, the computation is done closer to the sensors (\eg, on a gateway) and the application does not need to send all the information to the cloud to react to anomalous events. In this sense, the system is faster and more reliable with respect to the previous scenario.\\\\
As we write, Fog Computing does not stick to any universally-recognized standard architecture, while several research and development efforts have been focused on identifying the correct number of layers comprising the perfect Fog Computing architecture. More in detail, based on the implementation details, the literature boasts of architectures composed of three \cite{Bonomi2012,Nadeem2016,Taneja2016}, four \cite{Arkian2017}, five \cite{Dastjerdi2016}, six \cite{Aazam2015a}, and even eight \cite{Naha2018} layers. Arkian et al.\cite{Arkian2017} proposed a four-layer architecture, where the first three layers are those initially proposed in \cite{Bonomi2012}, while the fourth one is a vertical layer, namely the Data Consumer layer, used to make requests to the other layers. Dastjerdi et al. \cite{Dastjerdi2016} introduced a five-layer stack, where the IoT applications and the Software-defined resource management layers are located on top of the three traditional ones. Aazam et al. \cite{Aazam2015a} defined a six-layer structure by highlighting the functionality that should be implemented in the Fog. Recently, Naha et al. \cite{Naha2018} described a detailed and fine-grained architecture, where components are divided into eight different groups based on their functionality that defines the layer. Specifically: physical (sensors, actuators), Fog device (configuration and connectivity), Monitoring, Pre- and Post-Processing, Storage, Resource Management (resource allocation, scalability, reliability), Security (encryption/decryption, privacy, authentication), and Application layers. Another approach to Fog computing has been recently proposed by Sinaeepourfard et al. \cite{Sinaeepourfard2019} that use Fog computing as building block for a distributed-to-centralized data manager for smart cities. They identify three main layers: the Fog layer that contains IoT devices and performs some on-site processing. The Cloudlet layer that is the mid-layer, it is located in the same city of the fog layer and it is used as a communication layer among the different and distributed entities in the fog layer. The third layer is the (traditional) Cloud layer.
Notwithstanding this plethora of system architectures essentially reveals a lack of consensus among researchers, software architects, system integrators and IoT practitioners in general on a unified reference architecture for Fog Computing, it is important to take stock of the current, most mature, implementations from standardization, commercial, and open-source communities perspectives. Indeed, this is what the remaining of this paper will be all about.
\section{Standard Initiatives}
\label{sec:standards}
Some big names of industry and academia have already joined efforts, forming consortia with the aim of formalizing possible architectures for Fog Computing. One of the main initiative in this respect is the OpenFog consortium, originally funded in November 2015 by ARM, Microsoft Corp., Intel, Cisco, Dell, and Princeton University. OpenFog Consortium defines Fog Computing as \textit{"a horizontal, system-level architecture that distributes computing, storage, control, and networking functions closer to the users along a cloud-to-thing continuum"} \cite{OpenFogArch}. This consortium formally aims to fill the gap \cite{Bonomi2012} present in the design of IoT applications that are built using a "cloud-only" architecture. More in details, OpenFog has identified some "pillars" to distinguish Fog Computing from Cloud Computing, namely: \emph{i)} low latency, deployments, and computations near the data-sources (\ie, IoT devices); \emph{ii)} avoid migration costs (\ie, bandwidth); \emph{iii)} local communications instead of communications with remote end-nodes; \emph{iv)} management, network configuration and measurement deployed in fog nodes; \emph{v)} support for telemetry and analytics that should be sent to a remote system for orchestration and additional analytics.
\begin{figure}[t!]
	\centering
	\includegraphics[width=8.5cm]{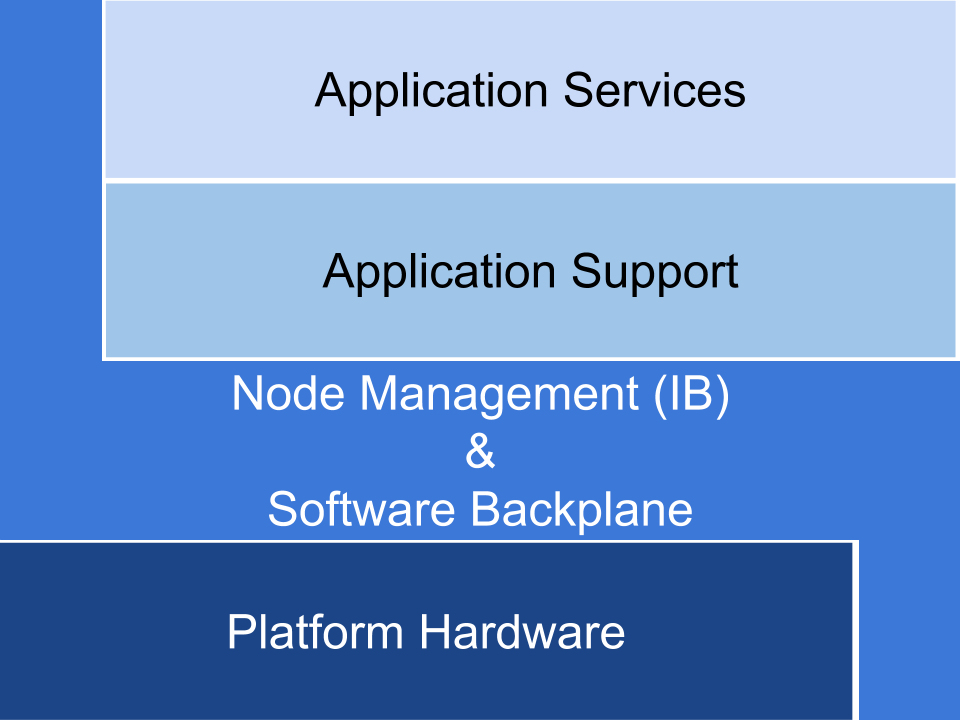}
	\caption{OpenFog Software architecture view (adapted from \cite{OpenFogArch}).}
	\label{fig:openfog-arch}
	\vspace{-0.5cm}
\end{figure}
The proposed architecture follows Bonomi's (Cisco)\cite{Bonomi2012} one: a three-layer stack where the Fog layer, composed of nodes called Fog Nodes, is split into four main sub-layers, namely Platform, Node Management, and Software Backplane, Application Support and Application Services (see \myfig{fig:openfog-arch}). 
The lower layer is the Platform Hardware that is the physical hardware of the Fog device.
The Node Management and Software Backplane layer is in charge of the general management of nodes and communications among end-points (\eg, remote cloud systems, edge devices, other Fog nodes).
The Application Support layer is a collection of micro-services that are not application-specific. These modules comprise databases, storage managers, networking stacks, security modules, message/event buses, runtime engines, analytic tools, etc.
The last module is the Application Services layer that offers many services to applications like Fog Connector services, Core services, Supporting services, Analytic services, Integration and User Interface services.

It is important to notice that this architecture has been conceived for quite powerful devices that are capable to offer both reactive and predictive capabilities. More in detail, reactive capabilities analyze the incoming data (\eg, vibrational signals) to discover if something is happening in the surrounding environment. This set of capabilities includes, for instance, Anomaly Detection, Rule Engines, Event processing, Sensor fusion and meta-sensors, supervisory control.
On the other hand, predictive capabilities, also referred to as forecasting capabilities, comprise Artificial Intelligence and Machine Learning based techniques that are able to identify patterns and to forecast future behaviors based on the incoming data. These predictive models may be directly inferred locally by the Fog node or, if required, in a hybrid fashion among different Fog nodes and Cloud systems. Since some operations over such models can be more demanding in terms of computational and memory power (\eg, model training) than others, these can be executed on more powerful remote cloud systems; after the training phase, such models are downloaded from the Cloud and deployed in the Fog nodes.\\

In August 2018, the Institute of Electrical and Electronics Engineers (IEEE) adopted the OpenFog Architecture as the reference architecture for Fog Computing for the IEEE~1934-2018 standard \cite{IEEEOpenFog}. This should allow developers and companies to build their own Fog-oriented applications using a standardized approach.  However, as we write, an open-source project implementing the whole IEEE~1934-2018 architecture, or that is at least fully compliant with all its specs, does not exist yet.

Then, in January 2019, the OpenFog Consortium and the Industrial Internet Consortium\footnote{\url{https://www.iiconsortium.org/}} (IIC) announced that they had finalized the agreement to join forces and merge the OpenFog consortium (and all of its working groups) under the umbrella of the IIC. This has been done since the two consortia were working on the common objectives and, in this way, they could boost the development and the deployment of Fog computing applications for the Industry 4.0 scenarios. However, this merge will not be an instantaneous process, as it will be finalized by 2020. Interestingly, as part of the agreement were the formal definitions of the terms Fog Computing and Edge Computing\footnote{\url{https://www.iiconsortium.org/IIC-OF-faq.htm}}. In more detail, Fog Computing, as defined by the OpenFog Consortium, is \textit{"a system-level horizontal architecture that distributes resources and services of computing, storage, control and networking anywhere along the continuum from Cloud to Things"}. On the other hand, the IIC defined Edge Computing as a \textit{"distributed computing that is performed near the edge, where the nearness is determined by the system requirements. The Edge is the boundary between the pertinent digital and physical entities, delineated by IoT devices"}. So, while the terms Edge and Fog Computing are often used interchangeably, the above definitions shed light on their conceptual differences. Specifically, Edge Computing leverages on processing resources \emph{already located} in the edge of the network (\ie, closer to end-users and IoT devices), while Fog Computing \emph{shifts} typical cloud capabilities towards the edge of the network, leveraging on the edge's resources (\eg, gateways, local servers, etc.), also facilitating the distribution of application logic in a Cloud-to-Thing continuum.

Also in January 2019, the Linux Foundation has started a new initiative, known as LF Edge\footnote{\url{https://www.lfedge.org}}. In this case, the declared objective is "\textit{to establish a unified open-source framework for the edge [...] contributing a new agnostic standard edge architecture}"\cite{LFEdgeLaunchPost}. This sub-foundation aims to create an open and interoperable ecosystem of software frameworks for fog/edge computing platforms, constrained to be vendor-neutral, hardware-independent, and technology-agnostic (\ie, cloud- and OS-independent). This would create a unified and aligned vision for the fog/edge computing paradigm by creating communities that will drive a better and more secure development of applications in the edge of the network. Currently, the LF Edge Foundation counts five different projects under its umbrella: 1) Akraino Edge Stack\footnote{\url{https://www.lfedge.org/projects/akraino/}}, which aims to create an open-source software stack to support cloud services on edge devices; 2) Edge Virtualization Engine\footnote{\url{https://www.lfedge.org/projects/eve/}} that aims to provide a complete, open-source, technology-agnostic, and standardized architecture to unify the methodology to design, develop, and orchestrate cloud applications over the edge; 3) Home Edge\footnote{\url{https://www.lfedge.org/projects/homeedge/}} that aims to develop a fog/edge computing framework for home automation, offering an open-source, robust, flexible and interoperable environment where devices can be simply integrated through a set of APIs, libraries, and runtimes; 4) The Open Glossary of Edge Computing\footnote{\url{https://github.com/lf-edge/glossary}}, that is an ambitious collaborative dictionary of expressions and terms related to the Fog/Edge Computing; 5) EdgeX Foundry\footnote{\url{https://www.edgexfoundry.org/}} that is an open-source and modular framework for Fog/Edge computing applications. EdgeX Foundry allows to plug modules and create custom functional blocks in a simplified way. Given its maturity, this project will be separately presented with more details in \mysec{sub-sec:fw-for-edge}.\\
The nature and the philosophy behind the LF Edge Foundation drive the open-source soul of this consortium. It is open to contributors, thus everyone can contribute to existing projects and can ask to incubate new ideas, while only the membership to the group requires the payment of an annual fee.\\

Other standardization initiatives are currently under development by different working groups in the Internet of Things ecosystem.\\
The Alliance for the Internet of Things Innovation (AIOTI) has proposed the AIOTI High Level Architecture (HLA) \footnote{\url{https://aioti.eu/wp-content/uploads/2018/06/AIOTI-HLA-R4.0.7.1-Final.pdf}}, with the aim to propose a deployment- and technology-agnostic architecture oriented to IoT applications. It is possible to host AIOTI HLA functionalities on top of Fog nodes. However, this architecture does not target Fog Computing but more in general IoT applications, thus we will not further discuss this initiative.\\
In 2012, the ITU-T has published the Recommendation Y.2060\footnote{\url{https://www.itu.int/rec/T-REC-Y.2060-201206-I}}, renumbered as Y.4000, that describes the general architecture for IoT applications. The document aims to define how an IoT application should be designed by defining a set of layers and corresponding capabilities. However, it assumes that the devices have simple and limited capabilities related to sensing, node management, and networking; a device may overlap with the gateway and, moreover, it assumes that there is a remote entity that manages and runs the IoT application.\\
The European Telecommunication Standard Institute (ETSI) has proposed the oneM2M Architecture\footnote{\url{https://www.etsi.org/deliver/etsi_ts/118100_118199/118101/02.10.00_60/ts_118101v021000p.pdf}}, that defines a software architecture for IoT and Machine-to-Machine (M2M) applications. This document splits the architecture into three different layers (Application Layer, Common Service Layer, and Network Service Layer), it defines many different types of node and their role in the architecture. It is possible to map some functionalities along the Cloud-to-Thing continuum. Moreover, ETSI has recently published a technical report\footnote{\url{http://member.onem2m.org/Application/documentapp/downloadLatestRevision/?docId=26390}} in which they are evaluating possible changes of the oneM2M architectures to introduce the concepts of Fog and Edge Computing.
\section{Fog Computing Platforms}
\label{sec:implementations}
In addition to the effort paid to define a standard Fog Computing architecture, many communities and private companies are also actively involved in designing and developing full-fledged software platforms able to cope with the unique requirements of this challenging computing paradigm. In the remaining of this section, we introduce some of the main actors playing on this stage.
\subsection{Frameworks for Fog/Edge Computing}
\label{sub-sec:fw-for-edge}
This section introduces two frameworks that are currently used in many applications, namely Apache Edgent and EdgeX Foundry.\\
Apache Edgent\footnote{\url{https://edgent.apache.org}} is a Java-based framework for edge stream analytics incubated by the Apache Foundation. Mainly, it enables a dataflow-based programming model suitable for fog/edge devices. Moreover, it provides a lightweight micro-kernel run-time environment embeddable in several off-the-shelf gateways and on other constrained devices able to execute a Java Runtime Environment (JRE). It also supports local and real-time analytics on data streams from the surrounding environment, such as vehicles, appliances, equipment, and so on. More in details, a fog application can integrate Apache Edgent in the fog layer and the framework uses analytics (\eg, split, union, filters, windowing, aggregations, etc.) to identify which data have to be streamed from the edge of the network to another computing entity (\ie, a cloud endpoint). This reduces the overall network bandwidth (and the associated cost of transmission, especially high in IoT contexts) and storage needs, also guaranteeing faster feedback towards local devices. Here, a developer can easily decide how data streams are managed inside his application and which computations have to be applied to which data. However, Apache Edgent provides only a few more capabilities than a stream manager, thus it does not come with a complete architecture to design an entire application. It can be easily integrated using the provided SDK and it can communicate with the outside world using well-known protocols such as, for instance, MQTT.

A more complete and mature platform for Fog Computing is EdgeX Foundry. The latter is a project originally donated by Dell Technologies to the open-source community in October 2017 and since then hosted under the umbrella of the LF Edge Foundation. Currently, EdgeX Foundry is supported and developed by more than 50 members coming from academia and industry. Since this project is constrained to be vendor-neutral, it provides a software solution that is not tied to any specific hardware or software supplier. The project aims to accelerate the deployment of IoT solutions by creating a unified and plug-and-play ecosystem that relies on an interoperability framework. This framework is implemented through an OS- and hardware-agnostic software platform for Fog and Edge devices that allows developers and companies to design new interoperable applications by combining standard connectivity interfaces (\eg, Wi-Fi, Bluetooth, BLE, etc.), common software modules and proprietary extensions. Moreover, the project leaderboard aims to contribute to create a common standard for IoT interoperability and to create a certification programme for hardware and software components in order to guarantee compatibility and interoperability.\\
\begin{figure*}[t!]
	\centering
	\includegraphics[width=14cm]{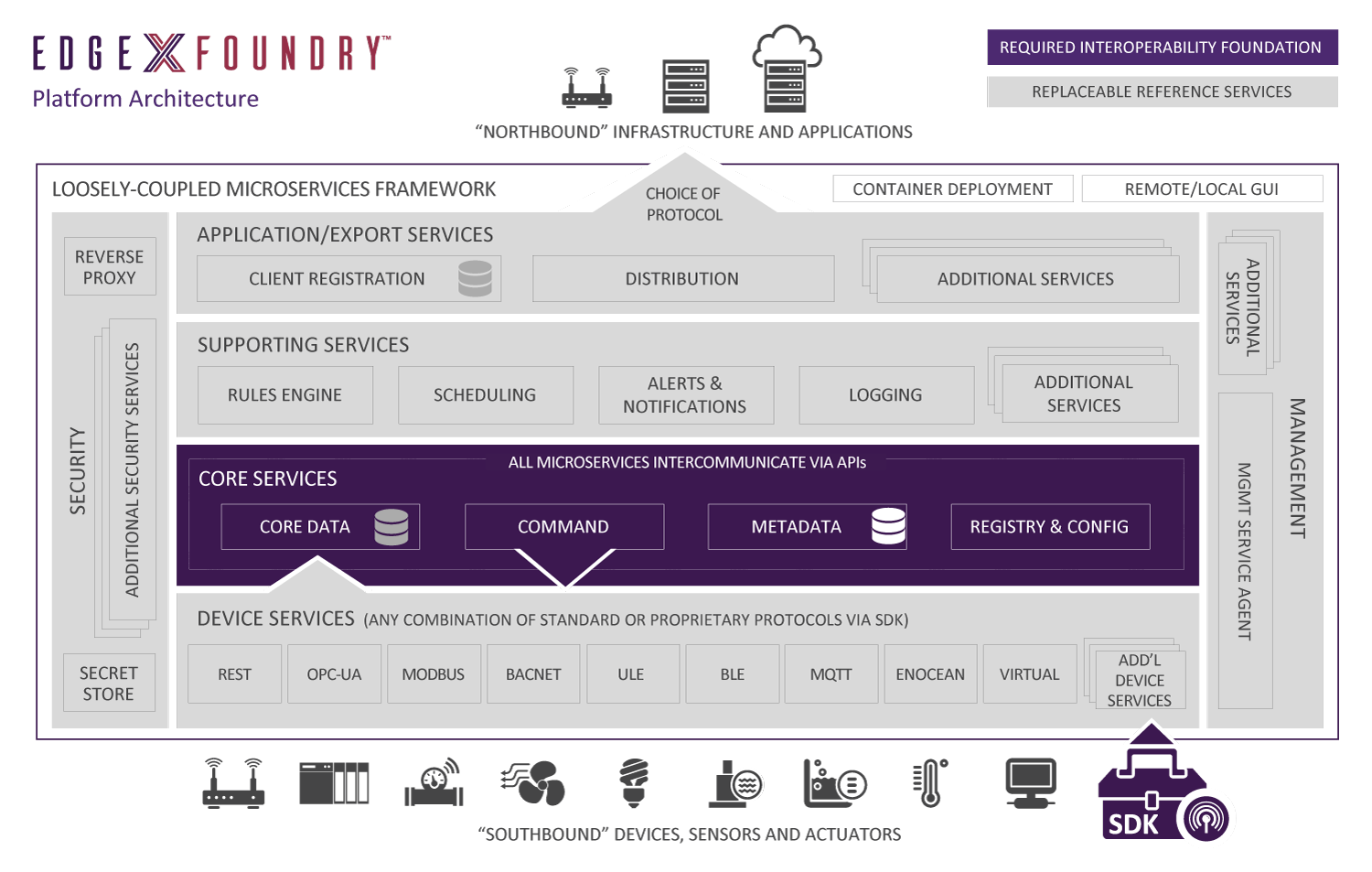}
	\caption{EdgeX Foundry platform architecture (taken from \cite{EdgeXFoundryArch}).}
	\label{fig:edgex-arch}
\end{figure*}
The EdgeX Foundry Architecture relies on the well-known three-layer Fog Architecture \cite{Bonomi2012}, as depicted in \myfig{fig:edgex-arch}, where edge devices and cloud systems are located below ("southbound", or edge layer) and above ("northbound", or cloud layer) the EdgeX Foundry software architecture, respectively. More in details, the EdgeX Foundry software architecture is located in the middle layer (\ie, fog layer), being composed of many sub-layers, as described in the following. Such architecture has been conceived starting from the micro-service paradigm \cite{Pautasso2017} that enables modular, scalable, secure, and technology-agnostic applications. Specifically, the chosen approach is the loosely-coupled micro-services architecture that requires a common layer to enable communications and data exchanges among modules using Inter-Processes Communications (IPC) APIs (\ie, REST APIs). This layer may be also distributed over more than one device if different services run on many Fog nodes. EdgeX Foundry can run on any edge/fog device such as gateways, routers, industrial PCs, servers, hubs, etc.\\
The EdgeX Foundry software architecture\cite{EdgeXFoundryArch}, which is located in the fog layer (see \myfig{fig:edgex-arch}), is composed of four horizontal sub-layers for the business logic definition, and two vertical sub-layers for security and management functionalities. The horizontal sub-layers are Device Services, Core Services, Supporting Services, and Application/Export Services. The Device Services layer comprises communication protocols and schema to interact with heterogeneous IoT Edge devices (\eg, BLE, MQTT, BACNET, MODBUS, REST, and so forth). It is also possible to integrate any missing protocol, by developing a micro-service that integrates such protocol through libraries. This allows companies to integrate their own proprietary technologies and protocols within EdgeX Foundry.
Core Services layer implements the Interoperability Foundation that is a set of micro-services (Core Data, Command, Metadata, Registry\&Configuration) required to build an application.
The Supporting Services layer exposes functionalities that are useful for all the applications defined at higher layers. It comprises Rule Engine, Scheduling Engine, Logging services an Alert\&Notification system. Similarly to protocols, it is possible to add new modules and services by developing a micro-service exposing the functionalities through an API.
The Application/Export Services layer contains the Client Registration and Distribution micro-services. The former allows clients to register for specific data streams inside the Fog node to receive updates when new data are available. The latter is used to register the fog node to other endpoints (\eg, MQTT, REST, etc.) to distribute data to other systems. Needless to say that even this layer can be extended, by developing new micro-services. Additional application micro-services run in this layer.
Finally, the two vertical sub-layers, the Security and the Management layers, expose micro-services that interact with all the four horizontal layers. The Management layer contains a micro-service to deploy new modules in the system and the Security module manages all the security operations like encryption, decryption, access, policies, and so on.\\
The EdgeX Foundry project and Architecture has been designed to create Edge/Fog applications for the Industrial IoT. Some of the main contributors (\eg, Dell) of this project have already commercialized some industrial computers that natively run EdgeX Foundry.
\subsection{Nebbiolo Technologies}
Commercially, some companies are offering products that implement the Fog Computing architecture. One of them is Nebbiolo Technologies that has been co-founded by Bonomi, \ie, one of the Fog Computing pioneers. Nebbiolo Technologies produces a complete Fog solution composed of nodes (\ie, fogNode\footnote{\url{https://www.nebbiolo.tech/wp-content/uploads/NFN-300-Datasheetv1.8FINAL-C-2018-Pantone.pdf}}), an operating system (\ie, fogOs\footnote{\url{https://www.nebbiolo.tech/wp-content/uploads/fogOS-Datasheetv1.5a-2018-Pantone.pdf}}), and a system manager (\ie, fogSM\footnote{\url{https://www.nebbiolo.tech/wp-content/uploads/NFN-300-Datasheetv1.8FINAL-C-2018-Pantone.pdf}}). Focusing on fogNodes, these are powerful machines built for industrial environments and equipped with powerful CPUs (Intel i5, i7 or Atom), solid-state disks (from 32 GB up to 512 GB), large RAM memory banks (from 4 GB to 16 GB), and networking interfaces (Ethernet, WiFi, and 3G/LTE). These machines support many functionalities and address many requirements of Fog Computing. However, these products are quite expensive and they are not as flexible as an open-source project could be, as all the code is developed by the company and it does not exist any community that supports and improves these products.
\subsection{Fog/Edge Computing-as-a-Service}
Another approach to Fog Computing is offered by Internet big players like Amazon Web Services (AWS), Microsoft, and so on. They are proposing solutions to implement Fog nodes with a cloud back-end and with support for computing capabilities like machine learning algorithms.\\
AWS offers Greengrass (AWS-GG)\cite{AWSggdg}, a software that extends the cloud capabilities of AWS Cloud closer to edge devices by directly enabling data collection and analysis on the edge of the network. Thus, developers can create, deploy and manage, via AWS-GG Cloud APIs, server-less pieces of code that typically run on cloud infrastructure, namely Lambda functions, on an edge device for local execution. Edge devices might be full-fledged computers, servers, virtual machines, but single-board mini-PC like Raspberry Pis. Lambda functions can be used to build IoT devices and they are triggered by events, messages from the cloud or other sources. These devices can communicate among them in a secure way, using authentication and authorization mechanisms, on the same network without any mediation with the remote cloud back-end. Running applications can continue their execution even in absence of connectivity. Furthermore, AWS-GG caches outbound and inbound messages using a local publish/subscribe message manager, based on MQTT, in order to preserve un-delivered messages. AWS-GG is composed of core software and an SDK to implement edge nodes, cloud APIs to manage devices and it supports many AWS products like Machine Learning Inference, Shadows implementation, group management, Lambda runtime, message manager, over-the-air updates, local resource access and so on.\\
AWS-GG organizes applications in groups that are collections of settings for AWS-GG core devices and devices that communicate with them (\ie, AWS IoT devices). Each group contains a list of Lambda Functions that can run on the core module, a list of MQTT subscriptions (message source, subject and target or destination) that enable communications among components, a list of devices that belong to the group, a list of resources and a core module that is in charge of execution and management of Lambda functions, local messaging among devices and the cloud.\\
\begin{figure}[t!]
	\includegraphics[width=9.0cm]{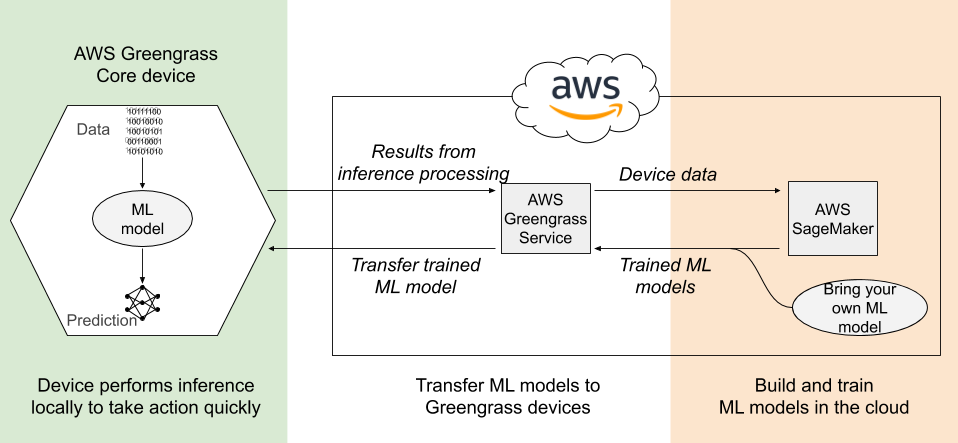}
	\caption{AWS Greengrass application architecture (adapted from \cite{AWSggdg}).}
	\label{fig:gg-ml-arch}
	\vspace{-0.5cm}
\end{figure}
AWS Greengrass offers the possibility to embed and infer machine learning models within a Lambda function running on an edge device. Model deployment is automatically performed by Greengrass. This approach (\myfig{fig:gg-ml-arch}) enables more intelligent edge applications that can reduce latency, costs (\ie, bandwidth and energy), and exploit powerful cloud systems to train models. Models can be built and trained using AWS SageMaker\footnote{\url{https://aws.amazon.com/sagemaker/}}, a cloud service developed to train deep-learning models using common frameworks like Tensorflow\footnote{\url{https://www.tensorflow.org/}}, Keras\footnote{\url{https://keras.io/}}, PyTorch\footnote{\url{https://pytorch.org/}}, Caffe2\footnote{\url{https://caffe2.ai/}}, and so on. Furthermore, Lambda functions, which infer models, can forward back incoming data to AWS S3\footnote{\url{https://aws.amazon.com/s3/}} (AWS cloud storage service) to provide new data samples to update or re-train models.\\
\begin{figure}[t!]
	\includegraphics[width=9.0cm]{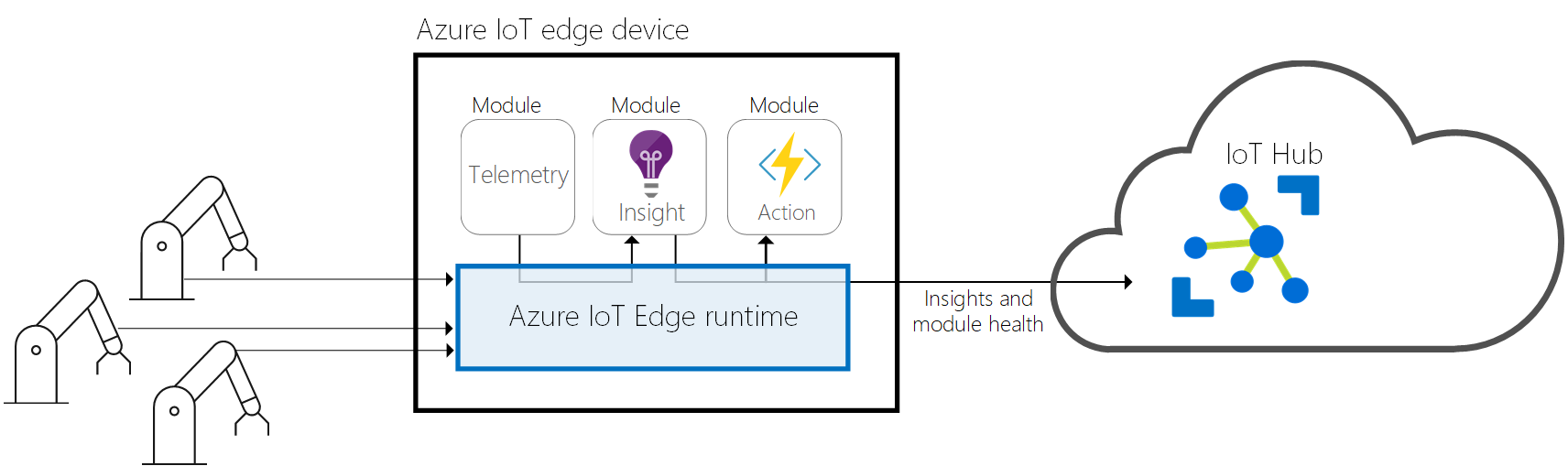}
	\caption{Azure IoT Edge application architecture (taken from \cite{AzureIoTEdge}).}
	\label{fig:azure-iot-edge-arch}
	\vspace{-0.5cm}
\end{figure}
Another suite of products for Fog Computing has been developed by Microsoft, namely Azure IoT Edge \cite{AzureIoTEdge}. As AWS Greengrass, it allows developers to deploy their own business logic closer to data sources in order to reduce latencies, bandwidth, and increase reactiveness. \myfig{fig:azure-iot-edge-arch} depicts the typical architecture of an Azure IoT Edge application. Azure IoT Edge is composed of three main components: containers that locally run Azure, user, or 3rd parties modules; Azure IoT Edge runtime that is executed on each device and manages all the other modules; a cloud-based interface to remotely monitor and manage a fleet of Azure IoT Edge devices. Azure IoT Edge runs on many different hardware architectures including Raspberry Pi, full-fledged computers, industrial computers, servers and so on. It has been designed by following the micro-service architecture \cite{Pautasso2017} to implement modules, which are Docker-compatible containers. Modules can be interconnected through pipelines to exchange data. Additional modules may add new classes of supported devices by exploiting the edge node's networking interfaces. Moreover, it enables machine learning applications by combining user-developed containers to run custom code, \ie, Feature Extraction, and Azure Machine Learning.\\
Concurrently, Microsoft is also developing a new set of tools for data analytics in the edge of the network known as Azure SQL Database Edge\footnote{\url{https://azure.microsoft.com/en-us/services/sql-database-edge/}}. As we write, this toolkit, only available for selected developers, enables local data processing on devices that support containerization, \eg, Linux- or Windows-based systems. It offers the possibility to execute graph-based local analytics, data-stream and time-series processing (\eg, data filtering, windowing, aggregation, etc.) before forwarding information to another entity, like a cloud endpoint, optimizing bandwidth and money. Moreover, it can locally execute in-database machine learning algorithms to identify patterns, anomalies, classify instances and so on. The model might be trained in the cloud and then locally deployed to be able to execute it offline and to minimize the inference latency.
\section{Comparisons and Considerations}
\label{sec:fog-comparisons}
All of the described solutions deal with the Fog Computing paradigm with very specific approaches, sometimes offering peculiar services and/or capabilities and hitting the market with different sales models and prices.

On the one side, Internet giants like Amazon, Microsoft, and Google provide cloud-assisted digital products targeting vertical domains like industry, home automation, buildings, vehicles, and so on. Essentially, they are comfortably playing the role of enablers for several novel applications within this framework, given the unique degree of integration with other cloud services they are already offering to their millions of users. However, Fog/Edge Computing has also the clear mandate to embed artificial intelligence closer to the data sources, meaning (among the other things) enabling the integration of some machine learning services and modules inside tinier IoT devices. On the contrary, these companies' products are not always available for typical embedded architectures (\eg, ARM-based boards) and sometimes they are not even publicly available yet. Moreover, even if such products are in general very flexible, they are also closed-source and often based on a cloud-centric approach. This means that the management and the orchestration of users' resources, models, and data are performed at cloud side. As said, this approach suffers from major issues, such as reliability, privacy, bandwidth, and so forth. Since these frameworks depend on a cloud back-end, the system may have unpredictable behavior if connectivity fails. Furthermore, data coming from devices are usually streamed to cloud endpoints to build a historical database and to update their predictive models. The amount of data might be too high to be streamed by the producers and/or stored by the cloud service (because of its volume, cost of the service, etc.). Typically, the billing of these products is based on batches of messages or per bunch of data. Finally, depending on the application, data might be sensitive, hence it should not be streamed.

On the other side, solutions like those offered by EdgeX Foundry and Nebbiolo tackle Fog Computing with a different approach: in principle, applications do not need to rely on any cloud infrastructure or remote service to implement their functionalities, but they can occasionally use cloud-based support to more efficiently tackle specific operations (usually the most demanding ones, in terms of computational and memory power). In this way, fog/edge devices may be empowered with high-performance computing units and big storage devices that allow applications where computation is pushed as much as possible closer to the edge of the network. For example, it is possible to train and execute machine learning and artificial intelligence algorithms directly on the edge. The developers can thus create modular applications keeping full control on the entire data-flow processing and of the devices. The typical approach is to exploit the micro-service paradigm \cite{Pautasso2017} as reference methodology. 

Concluding, this latter type of solutions are most suitable to meet requirements like \emph{i)} low latency and reactiveness (\eg, anomalies and faults are detected as fast as possible); \emph{ii)} reliability and cloud-independence (\eg, the system is not dependent on any specific cloud endpoint or service provider); \emph{i)} guarantee privacy (\eg, sensitive data like machinery vibrational data or e-health data are processed only locally); \emph{i)} reduced bandwidth (\eg, it is to be considered unfeasible to stream all raw data generated by all sensing devices, thus aggregation, feature extraction, and fusion are tasks to be resolved locally within the edge of the network). It goes without saying that this approach initially has higher economic costs with respect to cloud-centric approaches. However, these costs pay off later when no periodical payments of fees based on data volumes or transactions will be required.
\section{Conclusion}
\label{sec:conclusion}
The rise of Fog Computing is remarkably changing the way IoT applications are conceived and deployed along the Cloud-to-Thing continuum. In this paper, we have provided an overview of this novel computing paradigm from different perspectives. We have started our journey from the origins, trying to reconstruct the most pioneering steps made by the research community in this field. Then, we have dived into the main standardization initiatives, the most mature open-source solutions and the most advanced products/services already available on the market, in this latter case focusing on the Fog/Edge Computing paradigm offered with an as-a-Service model. We expect that Fog and Edge Computing will play a key role in the development of the IoT solutions of the future, mainly because of its by-design capabilities of enabling lower-latency, more secure, more cost-effective, and more complex applications, hence unleashing the true potential of the AI. 

%
%

\ifCLASSOPTIONcaptionsoff
  \newpage
\fi



%
\bibliographystyle{IEEEtran}

\begin{IEEEbiography}[{\includegraphics[width=1in,height=1.25in,clip,keepaspectratio]{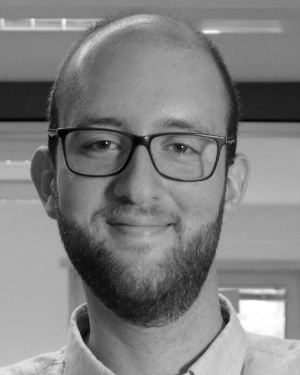}}]{Mattia Antonini} received the B.Sc. degree in Computer, Electronics and Telecommunication 
Engineering (Summa cum Laude) and the M.Sc. degree in Communication Engineering (Summa cum Laude) from the University of Parma, Parma, Italy, in 2014 and 2017, respectively. Currently, he is a Ph.D. student in Computer Science at the University of Trento, Trento, Italy and member of the OpenIoT Research unit (Open Platforms and Enabling Technologies for the Internet of Things) at FBK CREATE-NET, Trento, Italy.

His research interests include the design and development of open-source platforms for the Internet of Things with focus on machine learning-based applications in the Fog and Edge Computing. 
\end{IEEEbiography}

\begin{IEEEbiography}[{\includegraphics[width=1in,height=1.25in,clip,keepaspectratio]{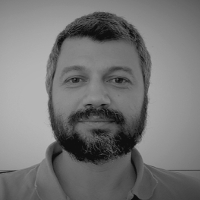}}]{Massimo Vecchio} received the M.Sc. degree in Information Engineering (Magna cum Laude) from the University of Pisa, Pisa, Italy and the Ph.D. degree in Computer Science and Engineering (with Doctor Europaeus mention) from IMT Institute for Advanced Studies, Lucca, Italy in 2005 and 2009, respectively. Starting from May 2015, he is an associate professor at eCampus University, while in September 2017 he has also joined FBK CREATE-NET, Trento, Italy to coordinate the research activities of the OpenIoT Research Unit. He is the project coordinator of AGILE (\url{www.agile-iot.eu}), a project co-founded by the Horizon 2020 programme of the European Union. His current research interests include computational intelligence and soft computing techniques, the Internet of Things paradigm and effective engineering design and solutions for constrained and embedded devices. Regarding his most recent editorial activity, he is a member of the editorial board of the Applied Soft Computing journal and of the newborn IEEE Internet of Things Magazine, besides being the managing editor of the IEEE IoT newsletters. 
\end{IEEEbiography}

\begin{IEEEbiography}[{\includegraphics[width=1in,height=1.25in,clip,keepaspectratio]{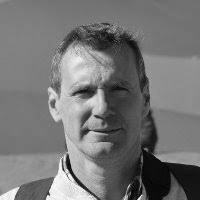}}]{Fabio Antonelli} is head of OpenIoT Research unit (Open Platforms and Enabling Technologies for the Internet of Things) at FBK CREATE-NET, Trento, Italy. With a Master's Degree in Electronics Engineering at Politecnico di Milano, Milan, Italy he worked for more than 15 years in the telco sector (within Alcatel and Telecom Italia groups) gaining extensive knowledge in experimental research, design, software development and management of ICT projects. More recently, in Fondazione Bruno Kessler, his interests have shifted on applied research in multimedia networking, architectures and platforms for the Internet of Things, where he has contributed and coordinated applied research activities in different European research projects in the Future Internet, Multimedia and Internet of Things domains.
\end{IEEEbiography}

\end{document}